\documentclass[12pt,preprint]{aastex}

\usepackage{graphics}

\newcommand{\be}{\begin{equation}}
\newcommand{\ee}{\end{equation}}

\slugcomment{submitted to : ApJ }

\shorttitle{Testing external Compton models} \shortauthors{Fan Z.H.}

\begin{document}

\title{A test on external Compton models for $\gamma$-ray active galactic nuclei}

\author{Zhonghui Fan\altaffilmark{1,2}, Xinwu Cao\altaffilmark{1}, Minfeng Gu\altaffilmark{1}}
\altaffiltext{1}{Shanghai Astronomical Observatory, Chinese
Academy
of Sciences, Shanghai 200030, China\\
E-mail: fanzh@shao.ac.cn; cxw@shao.ac.cn; gumf@shao.ac.cn}

\altaffiltext{2}{Graduate School of the Chinese Academy of Sciences,
BeiJing 100039, China }

\begin{abstract}
There is clear evidence that the $\gamma$-ray emission from active
galactic nuclei (AGNs) is attributed to the inverse Compton
scatterings in the relativistic blobs near the massive black holes.
If the soft seed photons are from the regions outside the blobs, a
linear relation between $(\nu F_{\nu,\gamma}/\nu F_{\nu, \rm syn}
u^{*})^{1/(1+\alpha)}$ and Doppler factor $\delta$ is expected,
where $\nu F_{\nu,\gamma}$ and $\nu F_{\nu, \rm syn}$ are
monochromatic $\gamma$-ray and synchrotron fluxes, respectively, and
$u^{*}$ is the energy density of soft seed photons \citep{D97}. We
estimate the soft photon energy density in the relativistic blobs
contributed by the broad line region (BLRs) in these $\gamma$-ray
AGNs using their broad-line emission data. The Doppler factors
$\delta$ are derived from their radio core and X-ray emission data,
based on the assumption that the X-ray emission is produced through
synchrotron self-Compton (SSC) scatterings. We find two nearly
linear correlations: $(\nu F_{\nu,\gamma}/\nu F_{\rm opt}
u^{*})^{1/(1+\alpha)} \propto \delta^{1.09}$, and $(\nu
F_{\nu,\gamma}/\nu F_{\rm IR} u^{*})^{1/(1+\alpha)} \propto
\delta^{1.20}$, which are roughly consistent with the linear
correlation predicted by the theoretical model for external Compton
scatterings. Our results imply that the soft seed photons are
dominantly from the BLRs in these $\gamma$-ray AGNs.
\end{abstract}


\section{Introduction}
All $\gamma$-ray AGNs are identified as flat-spectrum radio sources.
The third catalog of high-energy $\gamma$-ray sources detected by
the Energetic Gamma Ray Experiment Telescope (EGRET) on the Compton
Gamma Ray Observatory (CGRO) includes 66 high-confidence
identifications of blazars and 27 lower confidence potential blazar
identifications \citep{H99}. This provides a good sample for the
explorations on the radiative mechanisms of $\gamma$-rays from AGNs.
The violent variations in very short time-scales imply that the
$\gamma$-ray emission is closely related with the relativistic jets
in blazars. There are two kinds of models, namely, leptonic models
and hadronic models, proposed for $\gamma$-ray emission in blazars
(see Mukherjee 2001 for a review). According to the different
origins of the soft photons, the leptonic models can be classified
as two groups: synchrotron self-Compton (SSC) models and external
radiation Compton (EC) models (see Sikora $\&$ Madejski 2001 for a
review).

In the frame of SSC models, the synchrotron photons are both
produced and Compton up-scattered by the same population of
relativistic electrons in the jets of $\gamma$-ray blazars. The
synchrotron radiation is responsible for the low energy component in
radio bands, and the synchrotron photons are Compton up-scattered to
$\gamma$-ray photons by the same population of relativistic
electrons in the jets \citep{M92}. However, SSC models meet
difficulties with the observed rapidly variable fluxes in MeV-GeV
for some blazars. It has been realized that the processes other than
SSC may occur at least in some $\gamma$-ray blazars. One possibility
is soft seed photons being from the external radiation fields
outside the jets, namely, EC models. The origins of soft seed
photons may include the cosmic microwave background radiation, the
radiation of the accretion disk (including photons from the disk
scattered by surrounding gas and dust), infrared emission from the
dust or/and a putative molecular torus, and broad line region
(BLRs), etc \citep{D02}. Recently, \citet{S02} proposed that the
external radiation is from the BLRs for GeV $\gamma$-ray blazars
with flat $\gamma$-ray spectra, while the near-IR radiation from the
hot dust is responsible for MeV $\gamma$-ray blazars with steep
$\gamma$-ray spectra. In their model, the electrons are assumed to
be accelerated via a two-step process and their injection function
takes the form of a double power law with a break at the energy that
divides the regimes for two different electron acceleration
mechanisms.

The relations between $\gamma$-ray emission and emission in
different wavebands (such as optical, infrared and radio bands) may
provide clues on the radiative mechanism for $\gamma$-ray emission
(e.g., Zhou et al. 1997, Fan et al. 1998, Dondi \& Ghisellini 1995,
Zhang, Cheng \& Fan 2001, Yang \& Fan 2005). However, all these
correlation analysis cannot distinguish between the EC and SSC
models for $\gamma$-ray emission. \citet{D97} investigated the
$\gamma$-ray radiation from a homogeneous spherical blob
relativistically moving away from the central black hole. Their
calculations predicted a linear relation between $(\nu
F_{\nu,\gamma}/\nu F_{\nu, \rm syn} u^{*})^{1/(1+\alpha)}$ and
Doppler factor $\delta$, if the EC model is responsible for the
$\gamma$-ray radiation. Here $\nu F_{\nu,\gamma}$ and $\nu F_{\nu,
\rm syn}$ are monochromatic $\gamma$-ray and synchrotron fluxes,
respectively, $u^{*}$ is the energy density of soft seed photons,
and $\alpha$ is the photon energy spectral index of $\gamma$-rays.
For the SSC model, their calculations showed that $\nu
F_{\nu,\gamma}/\nu F_{\nu, \rm syn} u^{*}$ is independence of
$\delta$. \citet{HJ99} analyzed the correlations between the ratio
of $\gamma$-ray flux to the fluxes in different wavebands and
$\delta$ for a sample of EGRET AGNs. They found significant
correlations between $F_{\gamma}/\nu F_{\rm opt}$ and $\delta$, $
F_{\gamma}/\nu F_{\rm IR}$ and $\delta$. Their results suggested
that the EC model is responsible for $\gamma$-ray emission, though
the origin of the soft seed photons is still unclear.

Several different approaches are proposed to estimate the Doppler
factors $\delta$ of the jets in AGNs. \citet{G93} derived the
synchrotron self-Compton Doppler factor $\delta_{\rm SSC}$ of the
jets in AGNs from the VLBI core sizes and fluxes, and X-ray fluxes,
on the assumption of the X-ray emission being produced by the SSC
processes in the jets. \citet{GD96} assumed energy equipartition
between the particles and the magnetic fields in the jet components,
and derived the equipartition Doppler factor $\delta_{\rm eq}$. The
variability Doppler factor $\delta_{\rm var}$ is derived on the
assumption that the associated variability brightness temperature of
total radio flux density flares are caused by the relativistic jets
\citep{L99}.

In this paper, we use the multi-waveband data of $\gamma$-ray AGNs
to explore the radiative mechanism of $\gamma$-rays for these AGNs.
The Doppler factors $\delta$ are derived from VLBI and X-ray data by
using Ghisellini et al.'s approach. The cosmology with $H_{0}=70 \rm
{~km ~s^ {-1}~Mpc^{-1}}$, $\rm \Omega_{M}=0.3$, and $\rm
\Omega_{\Lambda} = 0.7$ have been adopted throughout the paper.

\section{Model}
\subsection{External Compton model}
In this paper, we mainly follow the model proposed by \citet{D97}.
We briefly summarize their model here (see Dermer, Sturner \&
Schlickeiser 1997 for details). Assuming that the seed photons are
produced externally to the jet, one can predict a correlation
between the Doppler factor $\delta$ and the ratio of the
$\gamma$-ray flux to the synchrotron radiation flux,
${\nu}F_{\nu,\gamma}/{\nu}F_{\nu,\rm syn}$, as shown by Eq. (27) in
\citet{D97}:
\begin{equation} \label{rate}
\rho_{\rm C/syn}=\frac {\nu F_{\nu,\gamma}} {\nu F_{\nu,\rm syn}}
\approx \left(\frac{\epsilon_s \bar{\epsilon}^*}{\epsilon_{\rm C}
\epsilon_H}\right)^{\alpha-1} \frac{u_{i}^*}{u_{H}}
{\delta}^{1+\alpha}
\end{equation}
where $\alpha$ is the photon energy spectral index of $\gamma$-rays;
$u_{i}^*$ and $u_{H}$ denote the energy density of monochromatic
photons in the external target radiation field and the blob's
magnetic field ($u_{H}=B^2/8\pi$, where the magnetic field strength
is $B\equiv 4.414\times 10^{13}\epsilon_{H}$), respectively;
$\bar{\epsilon}^*$ is monochromatic photon energy in the stationary
frame, $\epsilon_s$ and $\epsilon_{C}$ are the synchrotron photon
energy and inverse Compton scattering photon energy in blob frame,
respectively. Assuming that
\begin{equation}\label{energy}
(\epsilon_s \bar{\epsilon}^*)/(\epsilon_{\rm C} \epsilon_{H})\simeq
1
\end{equation}
\citep{D97}, Eq. (\ref{rate}) then becomes
\begin{equation}\label{rate2}
\rho_{\rm C/syn} \approx  \frac{u_{i}^*}{u_{H}} {\delta}^{1+\alpha}.
\end{equation}
For the homogeneous SSC model, the flux ratio of the SSC spectral
power flux  to the synchrotron spectral power flux $\rho_{\rm
SSC/syn}$ is independence of Doppler factor $\delta$ (see Eq. (28)
in Dermer, Sturner \& Schlickeiser 1997):
\begin{equation} \label{rate3}
\rho_{\rm SSC/syn}=\frac {\nu F_{\nu,\gamma}} {\nu F_{\nu,\rm syn}}
= \frac{2}{3} (\sigma_{\rm T} n_{eo} r_b)
\left(\frac{\epsilon_s}{\epsilon_{\rm C}}\right)^{\alpha -1} {\rm
ln} \overline{\Sigma}_{\rm C}(\epsilon_{\rm C}) \propto n_{eo} ,
\end{equation}
where $\sigma_{\rm T}$ is the Thomson cross section, $r_{b}$ is the
radius of the blob, $n_{eo}$ is the normalization factor of the
number density of nonthermal electrons, and $\overline{\Sigma}_{\rm
C}$ is the transformed Compton-synchrotron logarithm (see Eq. (25)
in Dermer, Sturner \& Schlickeiser 1997)
\begin{equation}
\overline{\Sigma}_{\rm C} (\epsilon_C)\equiv \frac{{\rm min}
\{\delta/ [\epsilon_C(1+z)], \gamma^{2}_{2}\epsilon_H,
\epsilon_C(1+z)/ (\delta \gamma^{2}_{1})\}}{{\rm max}
[\gamma^{2}_{1} \epsilon_H, \epsilon_C(1+z)/(\delta
\gamma^{2}_{2})]},
\end{equation}
where $\gamma_1$ and $\gamma_2$ are the lower limit and upper limit
of electronic Lorentz factors in the blob.

\subsection{Soft seed photon energy density}
In the EC models, the energy density of seed photon fields in a blob
may be the summation of the cosmic microwave background radiation,
the radiation of the accretion disk (including photons from the disk
scattered by surrounding gas and dust), infrared emission from the
dust and/or a putative molecular torus, and the BLRs, etc.. In this
work, we only focus on the soft seed photons contributed by the
broad line emission.

In principle, all broad emission line fluxes are needed to calculate
the total soft seed photon energy in the blobs contributed by
different broad emission lines. Usually, the fluxes of only one or
several broad emission lines are available for most sources in our
sample due to the restriction of redshift. We use the line ratios
presented by \citet{F91}, in which the relative strength of
Ly$\alpha$ is taken as 100, to convert the flux of available lines
into the flux of H$_{\beta}$ line. \citet{CP97} added the
contribution from the flux of H$\alpha$ $F_{\rm H\alpha}$, with a
value of 77. This gives a total relative flux $<F_{\rm BLR}>=555.77$
(narrow lines are not included). Due to the significant beaming
effect in the optical continuum emission, we use the relation
between the BLR size and H$_{\beta}$ luminosity \citep{W04} to
estimate the BLR radius. We re-fit their data and obtain the
correlation between $L_{\rm H_\beta}$ and $R_{\rm BLR}$:
\begin{equation}
 {\rm log}~R_{\rm BLR}~({\rm light-days}) =(1.323\pm0.086) 
 +(0.667\pm0.101)\; {\rm log}~(L_{\rm H_\beta}/10^{42}~{\rm
 ergs~s^{-1}}),
\end{equation}
for the present cosmology adopted in this paper. We calculate the
total photon energy density of the soft seed photons in the blob
from BLRs as \citep{F04}
\begin{equation}\label{ublr}
u_{\rm BLR}^{*}=\sum u_{\rm i}^{*}\approx\frac{555.77}{22} \times
u_{\rm H\beta}^{*}.
\end{equation}
The energy density $u_{\rm H\beta}^{*}$ of H$_{\beta}$ emission line
in the blob is given by
\begin{equation}\label{uhb}
u_{\rm H\beta}^{*}={\frac{1}{c}}\left(\frac{d_{L}}{R_{\rm
BLR}}\right)^{2}F_{\rm H\beta},
\end{equation}
where $d_L$ is the luminosity distance, $F_{\rm H\beta}$ is the flux
of H$_{\beta}$ emission line, and Eq. (\ref{uhb}) is valid for the
blob near the central black hole.

\subsection{The Doppler factor}
In this work, the Doppler factors $\delta$ are derived from the VLBI
radio core and X-ray emission data based on the assumption that the
X-ray emission is produced through synchrotron self-Compton (SSC)
processes. In the case of a moving sphere ($p=3+\alpha$) (see
Ghisellini et al. 1993 for details), $\delta$ of the blob can be
given as:
\begin{equation}
\delta=f(\alpha) F_{c} \left[ {\frac{{\ln ({\nu}_{b}/{\nu}_{s})}
}{{\ F_{\rm X} { \theta_{d}}^{6+4\alpha} {{\nu}_{x}}^{\alpha}
{{\nu}_{s}}^{5+3\alpha}} }} \right]^{1 / {(4+2\alpha)}} (1+z)~,
\end{equation}
where $F_{c}$ is the VLBI core radio flux density (in Jy) observed
at $\nu_s$ (GHz), $\theta_{d}$ is the VLBI core size (in mas),
$F_{\rm X}$ is the X-ray flux density (in Jy) observed at
${\nu}_{x}$ (keV), ${\nu} _{b}$ is the synchrotron high frequency
cutoff (assumed to be $10^{14}$ Hz), and the function
$f(\alpha)=0.08{\alpha}+0.14$ (here $\alpha=0.75$ is assumed).

\section{The Sample}
We search the literature and collect all available data of
$\gamma$-ray AGNs in EGRET catalog III. This leads to 40 sources, of
which 34 sources are the high-confidence identification blazars and
6 sources are the lower confidence potential  blazar identifications
listed in \citet{H99}.

Table 1 lists the VLBI and X-ray observations: (1) IAU name; (2)
confidence (A is high-confidence identifications of blazars, a is
lower confidence identifications of potential blazars); (3) type
classification of the source (BL=BL Lac object, Q= quasar); (4)
redshift; (5) VLBI core size $\theta _d $ in mas; (6) core radio
flux density $F_c$ at frequency $\nu _s$; (7) observation frequency
$\nu _s$ in GHz; (8) reference for the VLBI data; (9) 1 keV X-ray
flux density $F_X$ in $\mu $Jy; (10) reference for the X-ray flux;
(11) calculated Doppler factor $\delta$. The classification of
sources is according to the 11th edition catalogue of quasars and
active nuclei \citep{V03}. For those sources which have
multi-frequency VLBI observations, the VLBI data at the highest
frequency are chosen.

Table 2 lists the multiwavelength flux data and the flux density of
H$_{\beta}$ emission line: (1) IAU name; (2) the photon energy
spectral index of $\gamma$-rays; (3) $\gamma $-ray flux above 100
MeV, $F_\gamma $, in unit of ${10} ^{-8}$ photon {cm}$^{-2}$
{s}$^{-1}$; (4) V-band optical flux density $F_{\rm opt}$ in mJy;
(5) reference for the optical flux density; (6) near infrared flux
density $F_{\rm IR}$ at 2.2 $\mu$m in mJy; (7) reference for the
infrared flux density at 2.2 $\mu$m; (8) flux density at 5 GHz; (9)
reference for the flux density at 5 GHz; (10) flux density of
H$_{\beta}$ emission line, in unit of $10^{-15}$ ergs cm$^{-2}$
s$^{-1}$; (11) reference for the flux density of H$_{\beta}$; (12)
the total soft photon energy density $u_{\rm BLR}^{*}$ in the blob
from BLRs, in units of ergs cm$^{-3}$. The maximal values of the
infrared flux data are chosen from the literature when more than one
data are available. For six sources, we adopt the infrared flux from
2MASS archive. For the sources without flux density of H$_{\beta}$
emission line, we convert the flux density of other emission lines
into the flux density of H$_{\beta}$ emission line, as described in
Sect. 2.2, which is used to estimate the BLR size.

\section{Results}
We use the relation between BLR size and H$_{\beta}$ line luminosity
to calculate the BLR radius. The total photon energy in the blob of
the soft seed photons from BLRs are then estimated from Eq. (6). The
Doppler factors $\delta$ are derived from radio core and X-ray
emission data, as described in $\S$ 2.3. Assuming that the seed
photons are from BLRs, we investigate the correlation between
Doppler factor $\delta$ and the ratio of the $\gamma$-ray flux to
the synchrotron radiation flux. All the observational data and
results are listed in Tables $1-4$. The derived synchrotron
self-Compton Doppler factors $\delta$ are listed in Column (11) of
Table 1. The derived total photon energy densities of the soft seed
photons in the blob's frame from BLRs $u_{\rm BLR}^{*}$ are listed
in Column (12) of Table 2.

The $\gamma$-ray emission from these sources is violently variable.
In our analysis, we use the maximal integrated $\gamma$-ray fluxes
$F_{\gamma}$ ($>100$ MeV) for each source \citep{H99}. The
monochromatic flux at 100 MeV, 1 GeV and 20 GeV (i.e. $\nu F_{\nu}$)
are estimated from these maximal integrated $\gamma$-ray fluxes by
assuming that the SED of $\gamma$-ray is characterized as a power
law $F_{\nu}\propto \nu^{-\alpha}$ \citep{H99}.

The results of our correlation analysis are shown in Tables 3 and 4.
In both Tables, Columns (1) and (2), list two variables used in
analysis. Column (3) lists the number of sources. Columns (4) and
(6) list the intercept and the slope of the fitted line using OLS
bisector method \citep{IF90}, respectively. Columns (5) and (7) are
the standard deviations of intercept and slope, respectively. Column
(8) lists the pearson correlation coefficient. Column (9) lists the
chance probability.

\begin{figure}
\plotone{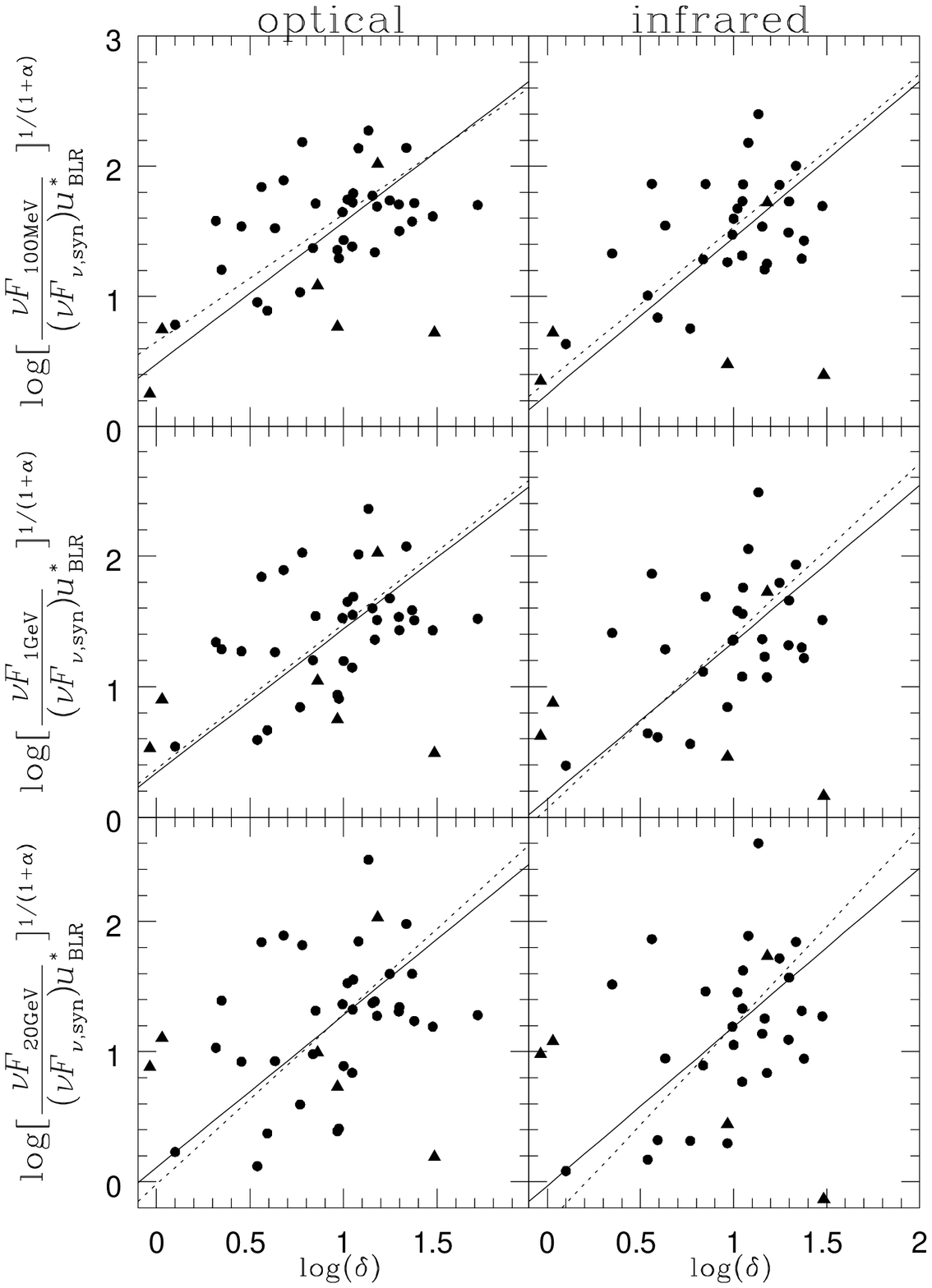} \caption{The relations of the Doppler factor
$\delta$ versus the flux ratio $(\nu F_{\nu,\gamma}/\nu F_{\nu, \rm
syn} u_{\rm BLR}^{*} )^{1/(1+\alpha)}$. The top, middle, and bottom
panels are the cases for $F_{\gamma}$ at 100 MeV, 1 GeV and 20 GeV,
respectively. While all left panels are for $F_{\rm syn}$ at 5500
{\AA}, all right panels are for $F_{\rm syn}$ at 2.2 $\mu$m. The
filled circles represent quasars, and the filled triangles are BL
Lac objects. \label{1}}
\end{figure}

The relations of the Doppler factor $\delta$ versus the flux ratio
$(\nu F_{\nu,\gamma}/\nu F_{\nu, \rm syn} u_{\rm BLR}^{*}
)^{1/(1+\alpha)}$ are plotted in Fig. \ref{1}. The top, middle, and
bottom panels in Fig. \ref{1} are the cases for $F_{\gamma}$ at 100
MeV, 1 GeV and 20 GeV, respectively. While all left panels are for
$F_{\rm syn}$ at 5500 {\AA}, all right panels are for $F_{\rm syn}$
at 2.2 $\mu$m. In Table 3, it can be seen that there are significant
correlations between $\delta$ and $(\nu F_{\nu, \gamma}/\nu F_{\rm
opt}u_{\rm BLR}^{*})^{1/(1+\alpha)}$, $\delta$ and $(\nu F_{\nu,
\gamma}/\nu F_{\rm IR}u_{\rm BLR}^{*})^{1/(1+\alpha)}$. Using OLS
bisector method, we obtain:
\begin{equation}\label{rateopt}
{\rm log}\left[\frac{\nu F_{\rm 100 MeV}}{(\nu F_{\rm opt})u_{\rm
BLR}^{*}}\right]^{\frac{1}{1+\alpha}} =(0.481\pm 0.105) + (1.089\pm
0.092)~ \rm {log}~\delta \; ,
\end{equation}
and
\begin{equation}\label{rateir}
{\rm log}\left[\frac{\nu F_{\rm 100 MeV}}{(\nu F_{\rm IR})u_{\rm
BLR}^{*}}\right]^{\frac{1}{1+\alpha}} =(0.253 \pm 0.128) + (1.203
\pm 0.104)~ \rm {log}~\delta  \; .
\end{equation}
The correlations are at a confidence level of 99.93 per cent for
optical flux at 5500 {\AA}, and at a confidence level of 99.55 per
cent for infrared flux at 2.2 $\mu$m, in the case of $\gamma$-ray
monochromatic flux at 100 MeV. When BL Lac objects are excluded, the
similar correlations are still present (see Table 3). The results
are similar for the cases of $\gamma$-ray at 1 GeV. For radio flux
at 5 GHz, we have not found any significant correlations between
$(\nu F_{\nu, \gamma}/\nu F_{\rm 5GHz}u_{\rm
BLR}^{*})^{1/(1+\alpha)}$ and $\delta$. For the average integrated
$\gamma$-ray flux, we also find the similar results, but the
correlations are slightly weaker compared with those for the maximum
integrated $\gamma$-ray flux.

In Fig. \ref{2}, we plot the relations between $F_{\gamma}/\nu
F_{\nu, \rm syn}$ and $\delta$ for different wavebands (optical, IR,
and radio), without considering the soft photon energy $u^*$, which
are similar to those as done by \citet{HJ99} (They used $(\nu
F_{\nu})_{\gamma}$ to represent the integrated $\gamma$-ray flux,
instead of the integrated $\gamma$-ray flux $F_{\gamma}$ used in
this work). We have not found any correlations between
$F_{\gamma}/\nu F_{\nu, \rm syn}$ and $\delta$ for each waveband
(see Table 4 for details), which are different from those given by
\citet{HJ99}.

\section{Discussion}
In most previous works, the size of the BLR is usually estimated
from the optical or UV continuum luminosity (e.g., Kaspi et al.
2000, Mclure \& Jarvis 2002, Vestergaard 2002). However, for
$\gamma$-ray blazars, the observed optical/UV continuum emission may
be dominantly from the relativistic jets, which is strongly beamed
to us. In this work, we estimate the size of the BLR from
H$_{\beta}$ luminosity instead of the optical or UV continuum
luminosity, and then the soft photon energy density in the blob.

\begin{figure}
\plotone{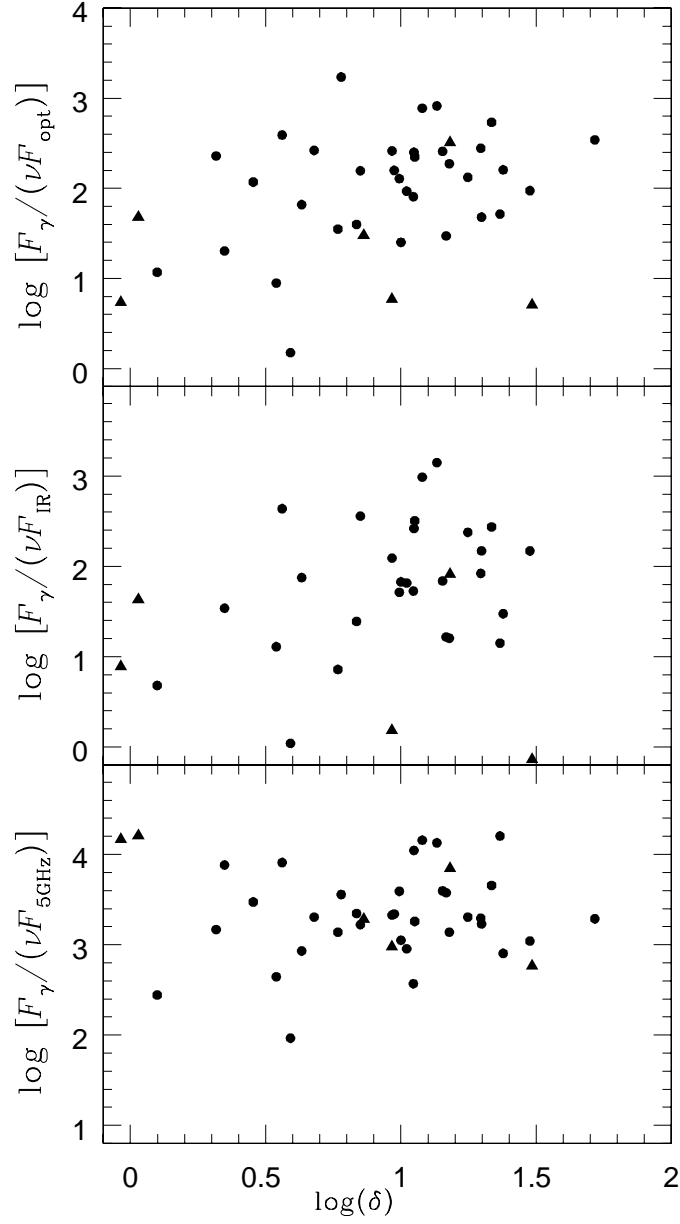} \caption{Doppler factor $\delta$ versus
$F_{\gamma}/\nu F_{\rm opt}$, $F_{\gamma}/\nu F_{\rm IR}$, and
$F_{\gamma}/\nu F_{\rm 5GHz}$ in top, middle, bottom panels,
respectively. The symbols are the same as in Fig. 1. \label{2}}
\end{figure}

We have found significant correlations $(\nu F_{\nu, \gamma} / \nu
F_{\rm opt}u_{\rm BLR}^{*})^{1/(1+\alpha)}$ -- $\delta$ and $(\nu
F_{\nu, \gamma}/\nu F_{\rm IR}u_{\rm BLR}^{*})^{1/(1+\alpha)}$ --
$\delta$, of which the slopes are close to unity. These are
consistent with the prediction of the EC model (see Eq. 3). For the
SSC model, the ratio of $\nu F_{\nu,\gamma}/\nu F_{\nu,\rm syn}$
should be independent of $\delta$ (see Eq. 4). Our results imply
that the $\gamma$-ray emission may be dominated by EC processes,
rather than SSC processes, at least for the $\gamma$-ray blazars in
this sample. On the origins of the external soft seed photons,
\citet{S94} proposed that the dominant contribution to the energy
density, as measured in the comoving frame of the radiating plasma,
comes from scattered or reprocessed portions of the centeal source
radiation, rather than from the direct radiation of the central
source. For an accretion disk, if a blob is sufficiently far from
the centeral engine of the AGN so that the accretion disk can be
approximated as a point source of photons, its photon energy density
(in the comoving frame) is $ u'_D \approx L_D/(4 \pi z^2
c\Gamma^2)$, where $L_D$ is the accretion disk luminosity, z is the
height of the blob above the accretion disk , and $\Gamma$ is the
Lorentz factor of the blob. Because of the blob leaves from the disk
with high velocity, soft photons from the disk are strongly
redshifted. For BLR, the reprocessed radiation is nearly isotropic
in the rest-frame of the central engine, it will be strongly
blue-shifted in the rest-frame of the relativistically moving blob.
Thus,
the ratio of the soft photon energy density measured in the rest
frame of the blob, $u^{' \rm BLR}_{\rm EC}/u^{'}_{D}\sim a_{\rm BLR}
(z/\langle r\rangle_{\rm BLR})^2 \Gamma^4 \gg 1$, where the fraction
of the radiation rescattered into the jet trajectory $a_{\rm BLR}
\sim 0.1$, and $\langle r\rangle_{\rm BLR}$ is the average distance
of the BLR from the centeral black hole \citep{B00}. For torus, the
typical thickness of the dust torus $H/R$ is around unity (e.g., Cao
2005), so that only a samll fraction of the photons from the disk
are scattered by the torus. The size of the torus is much larger
than that of the BLR. So, the soft photon energy density contributed
by the torus can be neglected compared with that from the BLR.
Although only the soft seed photons from the BLRs are considered in
this work, the results are in good agreement with the model
predictions, which implies that the soft photons are dominated from
the BLRs. \citet{HJ99} found significant correlations between
$F_{\gamma}/\nu F_{\nu,\rm syn}$ and $\delta$ for IR and optical
bands, while no correlation is present for our sample. There may be
two reasons, (1) the present sample consisting of 40 sources, which
is larger than theirs, (2) we adopt VLBI core sizes measured at
higher frequencies for some sources, and the derived Doppler factors
are therefore more reliable. We note that the results of
\citet{HJ99} implied that the equipartition parameter $\kappa_{\rm
eq}\equiv u_i^*/u_H$ remains constant between different sources. Our
results imply that the equipartition parameter $\kappa_{\rm eq}$ may
not be constant. This is may be one of reasons of the lack of
correlation in Fig. \ref{2}.

We note that Eq. (3) can be re-expressed as
\begin{equation}
{\rm log}[\nu F_{\gamma}/(\nu F_{\rm syn} u_{\rm
BLR})]^{1/(1+\alpha)} = A_1 + A_2 \; {\rm log}(B) + {\rm
log}(\delta),
\end{equation}
which can then be used to constrain the magnetic field strength B by
using Eq. (\ref{rateopt}) \& (\ref{rateir}). We can find $B\sim 1.7
- 2.8$ Gauss. Assuming $(\epsilon_s \bar{\epsilon}^{*})
/(\epsilon_{\rm C} \epsilon_{H})=1$, $\bar{\epsilon}^*$ is H$\beta$
emission line photon energy at 4861 {\AA} in the stationary frame,
the synchrotron photon energy $\epsilon_s$ is in optical-Infrared
region, $\gamma$-ray photon energy $\epsilon_{\rm C}$ is 100 MeV, we
find that the magnetic field strength $B$ in the blob frame is about
2 Gauss. We can see that the magnetic field strength derived from
equations (\ref{rateopt}) \& (\ref{rateir}) is agreement with the
model predictions. Moreover, these values of the magnetic field
strength are consistent with other researches (e.g., Maraschi \&
Tavecchio 2003). From $(\epsilon_s \bar{\epsilon}^*) /(\epsilon_{\rm
C} \epsilon_{H})=1$, we also can constrain the spectral region, in
which the synchrotron emission are radiated by the electron
population, responsible for $\gamma$-ray emission through inverse
Compton process. we find that the synchrotron photon frequency
corresponding to 100 MeV $\gamma$-ray photons are in
optical-Infrared region. This is in good agreement with the
significant correlations we found. However, for 5 GHz radio photons,
the corresponding photon energies via inverse Compton scattering may
be much lower than 100 MeV. In addition, the total 5 GHz radio flux
of these $\gamma$-ray sources are dominated by the radio core
emission, which is usually from optically thick region. These can
explain why there is no significant correlations between $(\nu
F_{\nu,\gamma}/\nu F_{\rm 5 GHz}u_{\rm BLR}^{*})^{1/(1+\alpha)}$ and
$\delta$.

In this work, we use SSC model to estimate $\delta$ assuming X-ray
is due to the SSC process. Our results show that $\gamma$-ray may
probably be produced by the EC process. Fitting the simultaneous
broadband spectrum of the FSRQ 3C279, \citet{H01} found that the SSC
model can fit the observed X-ray data very well, while the EC model
can fit to the $\gamma$-ray data successfully. For PKS 0528+134, a
similar conclusion was
drawn by \citet{M99}.
The BLR photons are mainly in optical/UV bands, and they are
strongly blueshifted to higher frequencies measured in the blob
frame with a factor of $\sim \Gamma$. So the observed energy of the
BLR photons scattered by the relativistic electrons in the blob
should at least in the range of $\sim \Gamma^{2} \gamma_{e,\rm
min}^{2} \epsilon_{\rm BLR}^{*}$, which should be much higher than 1
keV, for typical values $\Gamma \sim 10$, the lower limit of
electronic Lorentz factors $\gamma_{e,\rm min} \sim 100$ adopted
(e.g., Celotti \& Fabian 1993). Even for the photons from the disk
or torus, the energy of the scattered photons by the blobs is also
higher than 1 keV. So the X-ray emission at 1 keV should be produced
by SSC process.

No obvious statistic difference is found between BL Lac objects and
quasars in this sample (see Table 3), though the SSC model is
supposed to be responsible for $\gamma$-rays from BL Lac objects
(e.g. Dondi \& Ghisellini 1995), and the $\gamma$-ray varibility
properties of flat-spectrum quasars are different from those of BL
Lac objects \citep{V04}. There are only 6 $\gamma$-ray BL Lac
objects in our sample, and the remainder (10 BL Lacs) in the EGRET
AGNs catalogue are not included in this sample, because of the lack
of broad-line emission data for those BL Lac objects. All the BL Lac
objects in the present sample have relatively stronger broad lines
among the whole catalog of BL Lac objects. The BL Lac objects
included in this sample are more similar to quasars. These BL Lac
objects are somewhat special, which may be in the stage of the
transition from quasars to BL Lac objects (e.g. Cao 2003). For those
10 $\gamma$-ray BL Lac objects without broad emission lines, the SSC
mechanism may be important for their $\gamma$-ray emission, which is
beyond the scope of this paper.

\acknowledgements  We are thankful to the referee 
for insightful comments and constructive suggestions
that helped to improve this paper. This work is supported by the
National Science Foundation of China (grants 10325314, 10333020, and
10543002). This research has made use of the NASA/ IPAC
Extragalactic Database (NED), which is operated by the Jet
Propulsion Laboratory, California Institute of Technology, under
contract with the National Aeronautics and Space Administration.
This publication makes use of data products from the Two Micron All
Sky Survey, which is a joint project of the University of
Massachusetts and the Infrared Processing and Analysis
Center/California Institute of Technology, funded by the National
Aeronautics and Space Administration and the National Science
Foundation.

\clearpage

\begin{deluxetable}{ccccccccccc}
\tabletypesize{\scriptsize} \tablecaption{ VLBI and X-ray data of
the selected AGNs} \tablewidth{0pt} \tablehead{ \colhead{Source} &
\colhead{ID$^{*}$} & \colhead{type} & \colhead{z} &
\colhead{$\theta_{\rm d}$} & \colhead{$F_{c}(\nu_{s})$} &
\colhead{$\nu_s$} & \colhead{Ref.}  & \colhead{$F_{X}$} &
\colhead{Ref.} & \colhead{$\delta$}  \\
\colhead{} & \colhead{} & \colhead{} & \colhead{} & \colhead{mas}  &
\colhead{Jy} & \colhead{GHz} & \colhead{} & \colhead{$\mu$Jy} &
\colhead{} & \colhead{} } \startdata
0119$+$041 & a & Q & 0.637 & 0.310 & 0.580 &  8.55 & F00 &  0.12 & W94  &  2.08 \\
0208$-$512 & A & BL & 1.003 & 0.350 & 2.770 &  5.00 & HJ99&  0.61 & C97  & 15.17 \\
0234$+$285 & a & Q & 1.213 & 0.070 & 1.470 & 22.20 & J01 &  0.09 & C97  & 23.86 \\
0235$+$164 & A & Q & 0.940 & 0.080 & 0.705 & 43.20 & J01 &  0.78 & C97  &  2.23 \\
0336$-$019 & A & Q & 0.852 & 0.570 & 1.520 &  2.30 & HJ99&  0.10 & C97  & 13.58 \\
0414$-$189 & A & Q & 1.536 & 0.180 & 0.760 &  8.55 & F00 &  0.19 & S98  &  9.44 \\
0420$-$014 & A & Q & 0.914 & 0.060 & 2.724 & 43.20 & J01 &  0.44 & C97  & 15.10 \\
0454$-$234 & A & Q & 1.003 & 0.070 & 0.575 & 43.20 & J01 &  0.09 & C97  &  3.46 \\
0458$-$020 & A & Q & 2.286 & 0.024 & 0.934 & 43.20 & J01 &  0.10 & C97  & 52.14 \\
0521$-$365 & a & Q & 0.055 & 0.730 & 1.820 & 5.000 & HJ99&  2.12 & C97  &  1.26 \\
0537$-$441 & A & Q & 0.894 & 0.600 & 3.370 &  5.00 & HJ99&  0.81 & C97  &  6.86 \\
0804$+$499 & a & Q & 1.433 & 0.060 & 0.970 & 22.20 & J01 &  0.17 & C97  & 19.84 \\
0827$+$243 & A & Q & 0.939 & 0.050 & 1.407 & 43.20 & J01 &  0.34 & C97  & 11.16 \\
0836$+$710 & A & Q & 2.170 & 0.065 & 1.570 & 43.20 & J01 &  1.60 & C97  & 10.00 \\
0851$+$202 & A & BL & 0.306 & 0.043 & 1.640 & 43.20 & J01 &  0.97 & C97  &  9.27 \\
0954$+$658 & A & BL & 0.368 & 0.053 & 0.517 & 22.20 & J01 &  0.17 & C97  &  7.28 \\
1101$+$384 & A & BL & 0.030 & 0.240 & 0.366 &  5.00 & HJ99& 33.80 & C97  &  0.92 \\
1127$-$145 & a & Q & 1.187 & 0.140 & 1.060 & 22.20 & J01 &  0.34 & C97  &  4.29 \\
1156$+$295 & A & Q & 0.729 & 0.046 & 1.372 & 22.20 & J01 &  0.80 & C97  & 23.25 \\
1219$+$285 & A & BL & 0.102 & 0.090 & 0.263 & 22.20 & J01 &  0.41 & C97  &  1.07 \\
1222$+$216 & A & Q & 0.435 & 0.060 & 0.960 & 22.20 & J01 &  0.41 & C97  &  9.87 \\
1226$+$023 & A & Q & 0.158 & 0.135 & 8.040 & 43.20 & J01 & 12.30 & C97  &  3.91 \\
1243$-$072 & A & Q & 1.286 & 0.740 & 0.540 &  2.32 & F00 &  0.52 & S98  &  2.85 \\
1253$-$055 & A & Q & 0.538 & 0.072 & 5.440 & 43.20 & J01 &  1.34 & C97  & 14.69 \\
1334$-$127 & A & Q & 0.539 & 0.490 & 4.100 &  5.00 & SZ98 &  0.33 & S98  & 11.12 \\
1406$-$076 & A & Q & 1.494 & 0.075 & 0.833 & 22.20 & J01 &  0.18 & S98  & 12.00 \\
1424$-$418 & A & Q & 1.522 & 0.469 & 3.315 &  5.00 & FF0 &  0.18 & M05 & 17.67 \\
1504$-$166 & a & Q & 0.876 & 0.220 & 0.770 &  8.55 & F00 &  0.27 & S98  &  4.78 \\
1510$-$089 & A & Q & 0.360 & 0.056 & 1.458 & 43.20 & J01 &  0.74 & C97  &  5.85 \\
1611$+$343 & A & Q & 1.401 & 0.080 & 1.460 & 43.20 & J01 &  0.24 & C97  &  7.08 \\
1633$+$382 & A & Q & 1.814 & 0.075 & 1.553 & 22.20 & J01 &  0.42 & C97  & 21.64 \\
1730$-$130 & A & Q & 0.902 & 0.078 & 3.350 & 43.20 & J01 &  0.63 & C97  & 11.26 \\
1739$+$522 & A & Q & 1.375 & 0.073 & 0.973 & 22.20 & J01 &  0.16 & C97  & 14.25 \\
1741$-$038 & A & Q & 1.054 & 0.130 & 5.105 & 22.20 & J01 &  0.61 & C97  & 19.72 \\
1936$-$155 & A & Q & 1.657 & 0.680 & 0.860 &  2.32 & F00 &  0.05 & W94  &  9.26 \\
2200$+$420 & A & BL & 0.069 & 0.033 & 4.123 & 43.00 & L01 &  0.82 & G93 & 30.52 \\
2230$+$114 & A & Q & 1.037 & 0.040 & 2.429 & 43.20 & J01 &  0.29 & C97  & 30.02 \\
2251$+$158 & A & Q & 0.859 & 0.054 & 2.016 & 43.20 & J01 &  1.37 & C97  & 10.49 \\
2320$-$035 & A & Q & 1.410 & 0.190 & 0.330 &  8.55 & F00 &  0.17 & S98  &  3.64 \\
2351$+$456 & A & Q & 1.992 & 0.088 & 1.209 & 43.00 & L01 &  0.31 & G93 &  6.01 \\
\enddata
\tablenotetext{*}{A is high-confidence identifications of blazars, a
is lower confidence identifications of potential blazars}
\tablerefs{C97: \citet{C97}; F00: \citet{F00}; FF0: \citet{FF0};
G93: \citet{G93}; HJ99: \citet{HJ99}; J01: \citet{J01}; L01:
\citet{L01}; M05: \citet{M05}; SZ98: \citet{SZ98}; S98: \citet{S98};
W94: \citet{W94} }
\end{deluxetable}

\begin{deluxetable}{cccccccccccc}
\tabletypesize{\scriptsize} \tablecaption{ Multi-band data of the
selected AGNs} \tablewidth{0pt} \tablehead{ \colhead{Source} &
\colhead{$\alpha$} & \colhead{$F_{\gamma}$} & \colhead{$F_{\rm
opt}$} & \colhead{Ref.} & \colhead{$F_{\rm IR}$} &
\colhead{Ref.$^a$} & \colhead{$F_{\rm 5GHz}$} & \colhead{Ref.$^b$}
& \colhead{$F_{\rm H\beta}^c$} & \colhead{Ref.} & \colhead{${\rm log}\;u_{\rm BLR}^{*}$} \\
\colhead{} & \colhead{} & \colhead{} & \colhead{mJy} & \colhead{}  &
\colhead{mJy} & \colhead{} & \colhead{Jy} & \colhead{} & \colhead{}
& \colhead{} & \colhead{${\rm ergs\; cm^{-3}}$} } \startdata
0119$+$041 & 1.63 & 20.30 & 0.065 & K81 & ...   &     & 1.10 & FF0 & 5.800 & JB91   & $-$1.981 \\
0208$-$512 & 0.99 & 134.1 & 0.660 & D95 & 10.39 & 2MS & 3.31 & F98 & 11.65 & SF97   & $-$2.244 \\
0234$+$285 & 1.53 & 31.40 & 0.156 & D95 & 3.360 & B01 & 3.40 & FF0 & 19.66 & W84    & $-$2.388 \\
0235$+$164 & 0.85 & 65.10 & 6.600 & D95 & 15.48 & S96 & 1.90 & FF0 & 0.802 & C87    & $-$1.833 \\
0336$-$019 & 0.84 & 177.6 & 0.450 & C97 & 1.053 & B01 & 3.00 & FF0 & 12.00 & JB91   & $-$2.190 \\
0414$-$189 & 2.25 & 49.50 & 0.165 & K81 & ...   &     & 1.30 & FF0 & 0.388 & H78    & $-$1.903 \\
0420$-$014 & 1.44 & 64.02 & 0.296 & D95 & 13.90 & X98 & 4.40 & FF0 & 8.347 & SF97   & $-$2.162 \\
0454$-$234 & 2.14 & 14.70 & 0.910 & D95 & 2.510 & X98 & 2.00 & FF0 & 2.051 & S89    & $-$1.992 \\
0458$-$020 & 1.45 & 68.20 & 0.170 & D95 & ...   &     & 3.30 & FF0 & 0.185 & B89    & $-$1.938 \\
0521$-$365 & 1.63 & 31.90 & 2.000 & D95 & 19.50 & F93 & 9.20 & FF0 & 5.200 & SF97   & $-$1.172 \\
0537$-$441 & 1.41 & 91.10 & 2.050 & D95 & 13.29 & X98 & 4.00 & F98 & 3.818 & SF97   & $-$2.041 \\
0804$+$499 & 1.15 & 15.10 & 0.390 & D95 & 0.505 & C99 & 1.20 & FF0 & 1.935 & L96    & $-$2.112 \\
0827$+$243 & 1.42 & 111.0 & 0.391 & D95 & 1.497 & B01 & 0.97 & Z01 & 4.937 & G01    & $-$2.096 \\
0836$+$710 & 1.62 & 33.40 & 0.980 & D95 & 1.467 & B01 & 2.40 & FF0 & 13.83 & L96    & $-$2.545 \\
0851$+$202 & 1.03 & 15.80 & 4.000 & C97 & 61.80 & G85 & 2.70 & FF0 & 1.010 & S89    & $-$1.475 \\
0954$+$658 & 1.08 & 18.00 & 0.820 & C97 & ...   &     & 1.40 & FF0 & 0.417 & L96    & $-$1.409 \\
1101$+$384 & 0.57 & 27.10 & 17.80 & D95 & 50.13 & M90 & 0.72 & Z01 & 4.528 & M92    & $-$0.969 \\
1127$-$145 & 1.70 & 61.80 & 0.652 & C97 & 2.295 & B01 & 5.50 & FF0 & 29.26 & W95    & $-$2.438 \\
1156$+$295 & 0.98 & 163.2 & 5.100 & D95 & 74.80 & X98 & 1.80 & FF0 & 13.46 & W83    & $-$2.151 \\
1219$+$285 & 0.73 & 53.60 & 2.900 & C97 & 12.85 & X98 & 0.94 & Z01 & 0.049 & M96    & $-$0.683 \\
1222$+$216 & 1.28 & 48.10 & 0.390 & C97 & 3.873 & B01 & 1.40 & FF0 & 30.90 & S87    & $-$2.091 \\
1226$+$023 & 1.58 & 48.30 & 24.60 & D95 & 134.8 & X98 & 43.6 & FF0 & 1720. & K00    & $-$2.338 \\
1243$-$072 & 1.73 & 44.10 & 0.256 & K81 & ...   &     & 1.10 & FF0 & 6.838 & W86    & $-$2.256 \\
1253$-$055 & 0.96 & 267.3 & 15.10 & D95 & 108.8 & X98 & 13.0 & FF0 & 5.824 & SF97   & $-$1.922 \\
1334$-$127 & 1.62 & 20.20 & 0.185 & K81 & 1.124 & B01 & 4.40 & FF0 & 5.202 & S93    & $-$1.907 \\
1406$-$076 & 1.29 & 128.4 & 0.170 & D95 & 0.544 & 2MS & 1.00 & FF0 & 16.66 & W86    & $-$2.439 \\
1424$-$418 & 1.13 & 55.30 & 0.532 & K81 & 1.188 & 2MS & 3.80 & FF0 & 2.352 & SF97   & $-$2.162 \\
1504$-$166 & 1.00 & 33.20 & 0.197 & K81 & ...   &     & 2.80 & FF0 & 5.500 & SF97   & $-$2.087 \\
1510$-$089 & 1.47 & 49.40 & 1.180 & D95 & 23.09 & X98 & 3.30 & FF0 & 0.195 & JB91   & $-$1.291 \\
1611$+$343 & 1.42 & 68.90 & 0.390 & D95 & 0.681 & B94 & 4.00 & FF0 & 6.524 & W95    & $-$2.280 \\
1633$+$382 & 1.15 & 107.5 & 0.246 & D95 & 1.950 & X98 & 3.20 & FF0 & 9.965 & L96    & $-$2.434 \\
1730$-$130 & 1.23 & 104.8 & 0.520 & C97 & 1.457 & 2MS & 7.00 & FF0 & 7.000 & J84    & $-$2.132 \\
1739$+$522 & 1.42 & 44.90 & 0.155 & D95 & 2.314 & 2MS & 1.10 & FF0 & 4.186 & L96    & $-$2.209 \\
1741$-$038 & 1.42 & 48.70 & 0.155 & D95 & 2.074 & 2MS & 2.40 & FF0 & 2.109 & S89    & $-$2.014 \\
1936$-$155 & 2.45 & 55.00 & 0.105 & K81 & 0.886 & B01 & 1.40 & FF0 & 1.294 & SF97   & $-$2.105 \\
2200$+$420 & 1.60 & 39.90 & 5.900 & G93 & 165.0 & X98 & 5.60 & FF0 & 13.00 & V95    & $-$1.372 \\
2230$+$114 & 1.45 & 51.60 & 0.470 & D95 & 1.200 & X98 & 4.40 & FF0 & 14.24 & SF97   & $-$2.285 \\
2251$+$158 & 1.21 & 116.1 & 1.420 & D95 & 8.100 & X98 & 16.0 & FF0 & 47.03 & W95    & $-$2.391 \\
2320$-$035 & 1.00 & 38.20 & 0.153 & K81 & 0.551 & B01 & 0.80 & NED & 0.258 & B89    & $-$1.814 \\
2351$+$456 & 1.38 & 42.80 & 0.023 & G93 & ...   &     & 1.20 & FF0 & 3.691 & L96    & $-$2.323 \\
\enddata
\tablenotetext{a}{2MS is 2MASS} \tablenotetext{b}{NED is from NED
website} \tablenotetext{c}{the flux of H$_{\beta}$ emission line, in
unit of $10^{-15}$ ergs cm$^{-2}$ s$^{-1}$} \tablerefs{B89:
\citet{B89}; B94: \citet{B94}; B01: \citet{B01}; C87: \citet{C87};
C97: \citet{C97}; C99: \citet{C99}; D95: \citet{D95}; F93:
\citet{F93}; F98: \citet{F98}; FF0: \citet{FF0}; G85: \citet{G85};
G93: \citet{G93}; G01: \citet{G01}; H78: \citet{H78}; J84:
\citet{J84}; JB91: \citet{JB91};  K81: \citet{K81}; K00:
\citet{K00}; L96: \citet{L96}; M90: \citet{M90}; M92: \citet{M92};
M96: \citet{M96}; S87: \citet{S87}; S89: \citet{S89}; S93:
\citet{S93}; S96: \citet{S96}; SF97: \citet{SF97}; V95: \citet{V95};
W83: \citet{W83}; W84: \citet{W84}; W86: \citet{W86}; W95:
\citet{W95}; X98: \citet{X98}; Z01: \citet{Z01}}
\end{deluxetable}

\begin{deluxetable}{cccccccccc}
\tabletypesize{\scriptsize} \tablecaption{Results of the correlation
analysis} \tablewidth{0pt} \tablehead{ \colhead{y} & \colhead{x} &
\colhead{N} & \colhead{a} & \colhead{SD(a)} & \colhead{b} &
\colhead{SD(b)} & \colhead{r} & \colhead{prob} & \colhead{note} }
\startdata
log$\left[\frac{\nu F_{100 \rm MeV}}{(\nu F_{\rm opt})u_{\rm BLR}^{*}}\right]^{1/(1+\alpha)} $& log($\delta$)& 40 & 0.481 & 0.105 & 1.089 & 0.092 & 0.494 & 0.714D$-$03 &             \\
                                                                          &              & 34 & 0.650 & 0.106 & 0.979 & 0.105 & 0.457 & 0.461D$-$02 &excluding BLO\\
log$\left[\frac{\nu F_{1 \rm GeV}}{(\nu F_{\rm opt})u_{\rm BLR}^{*}}\right]^{1/(1+\alpha)}   $& log($\delta$)& 40 & 0.339 & 0.111 & 1.104 & 0.096 & 0.420 & 0.520D$-$02 &             \\
                                                                          &              & 34 & 0.367 & 0.134 & 1.113 & 0.127 & 0.432 & 0.785D$-$02 &excluding BLO\\
log$\left[\frac{\nu F_{20 \rm GeV}}{(\nu F_{\rm opt})u_{\rm BLR}^{*}}\right]^{1/(1+\alpha)}  $& log($\delta$)& 40 & 0.108 & 0.146 & 1.172 & 0.120 & 0.294 & 0.586D$-$01 &             \\
                                                                          &              & 34 &-0.020 & 0.190 & 1.307 & 0.169 & 0.388 & 0.186D$-$01 &excluding BLO\\
log$\left[\frac{\nu F_{100 \rm MeV}}{(\nu F_{\rm IR})u_{\rm BLR}^{*}}\right]^{1/(1+\alpha)}  $& log($\delta$)& 33 & 0.253 & 0.128 & 1.203 & 0.104 & 0.464 & 0.446D$-$02 &             \\
                                                                          &              & 28 & 0.346 & 0.150 & 1.181 & 0.142 & 0.484 & 0.600D$-$02 &excluding BLO\\
log$\left[\frac{\nu F_{1 \rm GeV}}{(\nu F_{\rm IR})u_{\rm BLR}^{*}}\right]^{1/(1+\alpha)}    $& log($\delta$)& 33 & 0.137 & 0.144 & 1.202 & 0.119 & 0.375 & 0.255D$-$01 &             \\
                                                                          &              & 28 & 0.065 & 0.193 & 1.319 & 0.179 & 0.464 & 0.912D$-$02 &excluding BLO\\
log$\left[\frac{\nu F_{20 \rm GeV}}{(\nu F_{\rm IR})u_{\rm BLR}^{*}}\right]^{1/(1+\alpha)}   $& log($\delta$)& 33 &-0.029 & 0.200 & 1.217 & 0.172 & 0.246 & 0.156D$+$00 &             \\
                                                                          &              & 28 &-0.321 & 0.281 & 1.521 & 0.251 & 0.424 & 0.186D$-$01 &excluding BLO\\
log$\left[\frac{\nu F_{100 \rm MeV}}{(\nu F_{\rm 5GHz})u_{\rm BLR}^{*}}\right]^{1/(1+\alpha)}$&log($\delta$)& 40 & 1.222 & 0.093 & 0.980 & 0.039 & 0.164 & 0.301D$+$00 &             \\
                                                                          &              & 34 & 1.147 & 0.103 & 1.030 & 0.089 & 0.342 & 0.404D$-$01 &excluding BLO\\
log$\left[\frac{\nu F_{1 \rm GeV}}{(\nu F_{\rm 5GHz})u_{\rm BLR}^{*}}\right]^{1/(1+\alpha)}  $& log($\delta$)& 40 & 1.050 & 0.136 & 1.027 & 0.081 & 0.071 & 0.657D$+$00 &             \\
                                                                          &              & 34 & 0.848 & 0.156 & 1.181 & 0.136 & 0.311 & 0.646D$-$01 &excluding BLO\\
log$\left[\frac{\nu F_{20 \rm GeV}}{(\nu F_{\rm 5GHz})u_{\rm BLR}^{*}}\right]^{1/(1+\alpha)} $& log($\delta$)& 40 & 2.753 & 0.232 &-1.002 & 0.193 & -.002 & 0.990D$+$00 &             \\
                                                                          &              & 34 & 0.487 & 0.257 & 1.347 & 0.232 & 0.282 & 0.957D$-$01 &excluding BLO\\
\enddata
\end{deluxetable}

\begin{deluxetable}{cccccccccc}
\tabletypesize{\scriptsize} \tablecaption{Results of the correlation
analysis} \tablewidth{0pt} \tablehead{ \colhead{y} & \colhead{x} &
\colhead{N} & \colhead{a} & \colhead{SD(a)} & \colhead{b} &
\colhead{SD(b)} & \colhead{r} & \colhead{prob} & \colhead{note} }
\startdata
log$\left[\frac{F_{\gamma}}{\nu F_{\rm opt}}\right] $ &log($\delta$) & 40 & 0.670 & 0.199 & 1.371 & 0.170 & 0.338 & 0.279D$-$01 &             \\
                                       &              & 34 & 0.687 & 0.269 & 1.423 & 0.226 & 0.368 & 0.266D$-$01 &excluding BLO\\
log$\left[\frac{F_{\gamma}}{\nu F_{\rm IR}}\right]  $ &log($\delta$) & 33 & 0.372 & 0.253 & 1.386 & 0.262 & 0.237 & 0.171D$+$00 &             \\
                                       &              & 28 & 0.151 & 0.352 & 1.698 & 0.324 & 0.382 & 0.366D$-$01 &excluding BLO\\
log$\left[\frac{F_{\gamma}}{\nu F_{\rm 5GHz}}\right]$ &log($\delta$) & 40 & 4.286 & 0.142 &-1.018 & 0.087 & -.039 & 0.808D$+$00 &             \\
                                       &              & 34 & 2.211 & 0.179 & 1.157 & 0.152 & 0.249 & 0.144D$+$00 &excluding BLO\\
\enddata
\end{deluxetable}

\end{document}